\documentclass[11pt]{article}
\usepackage{graphicx}
\usepackage{subfigure}
\usepackage{amsmath}
\usepackage{amsfonts}
\usepackage{cancel}

\begin{document}

\newcommand{\Ima}{\textrm{Im}}
\newcommand{\Rea}{\textrm{Re}}
\newcommand{\mev}{\textrm{ MeV}}
\newcommand{\be}{\begin{equation}}
\newcommand{\ee}{\end{equation}}
\newcommand{\ba}{\begin{eqnarray}}
\newcommand{\ea}{\end{eqnarray}}
\newcommand{\gev}{\textrm{ GeV}}
\newcommand{\nn}{{\nonumber}}
\newcommand{\dtres}{d^{\hspace{0.1mm} 3}\hspace{-0.5mm}}

\title{On the nature of the $K^*_2(1430)$, $K^*_3(1780)$, $K^*_4(2045)$, $K^*_5(2380)$ 
and $K^*_6$ as  $K^*$--multi-$\rho$ states
}

\author{J. Yamagata-Sekihara$^1$ L. Roca$^2$ and E. Oset$^1$ \\
{\small{\it $^1$Departamento de F\'{\i}sica Te\'orica and IFIC,
Centro Mixto Universidad de Valencia-CSIC,}}\\
{\small{\it Institutos de
Investigaci\'on de Paterna, Aptdo. 22085, 46071 Valencia, Spain}}\\
{\small{\it $^2$Departamento de F\'{\i}sica. Universidad de Murcia. E-30071, Murcia. Spain}}
}

\date{\today}

\maketitle

 \begin{abstract}
 
We show that the $K^*_2(1430)$, $K^*_3(1780)$, $K^*_4(2045)$,
$K^*_5(2380)$ and a not yet discovered $K_6^*$ 
resonance are basically molecules made of an increasing 
number
of $\rho(770)$ and one $K^*(892)$ mesons.
The idea relies on the fact that the
vector-vector interaction in s-wave
with spins aligned is very strong both for
$\rho\rho$ and $K^*\rho$.
We extend a recent work, where several resonances showed up as
multi-$\rho(770)$ molecules, to the strange sector including the
$K^*(892)$ into the system. The resonant structures show up in 
the multi-body scattering amplitudes, which are
evaluated in terms  of the unitary two-body vector-vector scattering amplitudes by using
the fixed center
approximation to the Faddeev equations.

\end{abstract}

\section{Introduction\label{sec:1}}

The nature and structure of hadronic resonances is a prime issue in the
understanding of the strong interaction. Besides the simplest
quark-antiquark picture, other structures
such as tetraquarks, glueballs or meson molecules may be 
dominant in the contribution to the wave
function 
of some mesonic resonances. 
Regarding the meson molecule contribution, important milestones have
been reached by 
the unitary extensions of chiral perturbation theory (UChPT), the chiral unitary
approach. With the only input of lowest orders chiral Lagrangians and 
 the implementation of unitarity in coupled channels, many 
resonances can be interpreted as meson-meson or meson-baryon
molecules
\cite{Kaiser:1995cy,npa,iam,nsd,Kaiser:1998fi,angels,juanenrique,ollerulf,carmenjuan,hyodo},
which are usually called dynamically generated resonances.

The use of vector mesons as building blocks
within the chiral unitary approach has been recently considered
both in the interaction of vector mesons with
baryons \cite{sourav,angelsvec} and among themselves
\cite{Molina:2008jw,gengvec,nagahiro},
  starting
from a lowest order hidden gauge symmetry Lagrangian
\cite{hidden1,hidden2,hidden3,hidden4}.
One of the main outputs of refs.~\cite{Molina:2008jw,gengvec} was the
very strong attraction of the s-wave vector-vector interaction with spin 2, to the point
of generating dynamically the $f_2(1270)$ and the $K^*_2(1430)$ resonances, among
others, as
$\rho\rho$ and $K^*\rho$ molecules respectively with a very strong
binding energy.
Due to this strong attraction, in ref.~\cite{Roca:2010tf}
the issue of whether is it possible to obtain bound systems 
with increasing  number of $\rho$
mesons as building blocks was addressed. Indeed, it was
found in ref.~\cite{Roca:2010tf} that  the resonances 
$\rho_3(1690)$ ($3^{--}$),   $f_4(2050)$ ($4^{++}$), $\rho_5(2350)$
($5^{--}$) and   $f_6(2510)$ ($6^{++}$) are basically molecules of
increasing number of $\rho(770)$ particles. The multi-body
interaction was written in terms  of the two-body scattering amplitudes, 
using the fixed center
approximation of the Faddeev equations (FCA) without the inclusion of any new
free parameters.

The main aim of the present work is to extend the analysis of
ref.~\cite{Roca:2010tf} to the strange sector including the interaction
of the $K^*(892)$ resonance with several $\rho(770)$ mesons since, as mention above, 
the $\rho K^*$ interaction 
 in s-wave and spin 2 is strongly attractive. Actually, it is strong enough
to generate the $K_2^*(1430)$
resonance as a $K^*\rho$ quasibound state or molecule. Thus the main 
aim in the present work is, 
first, study the interaction of one $K^*$
and two $\rho$ mesons (the latter ones 
clustered in an $f_2(1270)$) in order to see if some
resonance structure shows up in the multi-body scattering amplitude.
If this is the case, it could correspond to the $K^*_3(1780)$ resonance.
The procedure can be naturally extended including more $\rho$ mesons
 into the system.
Indeed, a $\rho$ meson can be added to this generated $K^*_3(1780)$ 
to try to
see if in this four-body scattering amplitude there is evidence of a
 resonant structure that could be associated to the 
$K^*_4(2045)$, (it could also be found in 
$f_2(1270)$-$K_2^*(1430)$ interaction, which we will also study). Similarly, adding more $\rho$ mesons
we will study systems of one $K^*$ and up to five $\rho$ mesons to see, in
the end, if the $K^*_2(1430)~(2^{+})$, $K^*_3(1780)~(3^{-})$, $K^*_4(2045)~(4^{+})$,
$K^*_5(2380)~(5^{-})$ and a possible $K^*_6~(6^{+})$ resonances (the latter one 
not yet experimentally observed)
are basically molecules made of an increasing 
number
of $\rho(770)$ and one $K^*(892)$ mesons.

\section{Two-body interaction: vector-vector unitarization\label{sec:2}}

For the evaluation of the multi-body interactions that we are going 
to consider in the
present work, one of the main ingredients is
the two-body $\rho\rho$ and $K^*\rho$ unitarized
scattering amplitudes. The unitarized $\rho\rho$ amplitude is explained
in detail in ref.~\cite{Molina:2008jw} and summarized
 in ref.~\cite{Roca:2010tf}. In the latter work 
the extension to evaluate the multi-$\rho$
scattering amplitudes was also explained. The main novelty of the present work with respect
to ref.~\cite{Roca:2010tf}
is the inclusion of strangeness via the $K^*$ resonance. For completeness,
we briefly summarize here
the model of \cite{gengvec} for the $K^*\rho$ interaction.
(We refer to \cite{gengvec} for further details.)

Following the ideas of the chiral unitary approach,
the implementation of unitarity in the  scattering amplitudes
leads to the full two-body scattering amplitude for a given partial wave,
which can be  written by means of the
Bethe-Salpeter equation in coupled channels:

\be
t=V+VGt=(1-VG)^{-1}V
\label{eq:bethesalpeter}
\ee
where the kernel $V$ is a matrix containing the tree-level 
vector-vector transition 
amplitudes (kernel of the Bethe-Salpeter equation) and $G$
is a diagonal matrix which elements are the
 loop function for two vector mesons:
\begin{equation}\label{eq:Gsharp}
G=i\int\frac{d^4q}{(2\pi)^4}\frac{1}{q^2-M^2_{V1}}\frac{1}{q^2-M^2_{V2}},
\end{equation}
where $M_{V1}$ and $M_{V2}$ are the masses of the two vector-mesons
of the corresponding channel.
The widths of the vector mesons inside the loop $G$ are taken into
account by folding Eq.~(\ref{eq:Gsharp}) with their spectral
functions as explained in ref.~\cite{gengvec}.
The loop function in Eq.~(\ref{eq:Gsharp}) can be regularized either
with a three-momentum cutoff or with dimensional regularization. The
equivalence of both methods for meson-meson scattering 
 was shown in ref.~\cite{iam}.
In ref.~\cite{gengvec} the regularization method was used 
with subtraction constants $a_{\rho\rho}=-1.636$,
and $a_{\rho K^*}=-1.85$, chosen to get the right mass of the 
$f_2(1270)$ and the
$K^*_2(1430)$ respectively. However, moderate changes in these
parameters
only produce small quantitative differences in the 
results \cite{gengvec}. The use of dimensional regularization with
the subtraction constants given above, is equivalent to considering
the cutoff method
with a three-momentum cutoff of $875\mev$ for $\rho\rho$
and $1123\mev$ for $\rho K^*$.

\begin{figure}[!t]
\begin{center}
\includegraphics[width=0.8\textwidth]{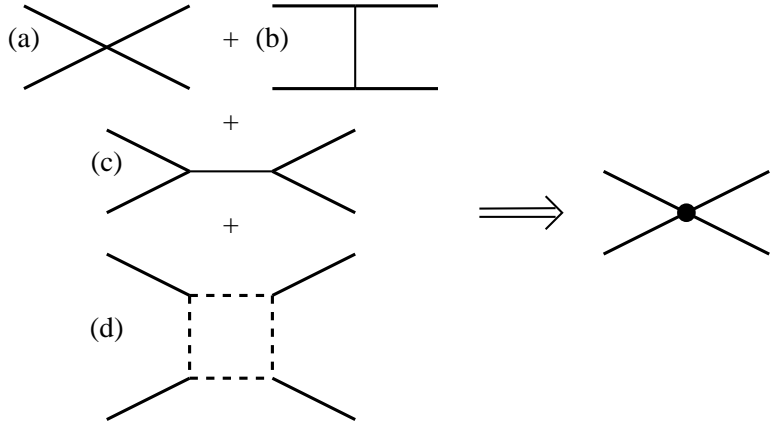}
\caption{Mechanisms contributing to the kernel $V$ (thick dot) of the
Bethe-Salpeter equation, Eq.~(\ref{eq:bethesalpeter}), for 
vector-vector scattering. Solid lines represent vector mesons and dashed
 lines pseudoscalar ones.}
\label{fig:vvdiagram1}
\end{center}
\end{figure}

The contributions to the vector-vector potential $V$ needed in
Eq.~(\ref{eq:bethesalpeter}), represented by a thick dot in Fig.~\ref{fig:vvdiagram1}, are depicted by the diagrams 
at the left of Fig.~\ref{fig:vvdiagram1}. In this figure
  the solid lines represent vector mesons and the dashed
lines pseudoscalar ones.
The vertices needed to evaluate these diagrams 
are obtained from the hidden gauge symmetry Lagrangian
\cite{hidden1,hidden2,hidden3,hidden4} for vector mesons. Actually we need the 4-vectors,
3-vectors and one vector--2-pseudoscalars contact terms. Explicit expressions
for the Lagrangians and the $V-$matrix elements
 can be found 
in refs.~\cite{Molina:2008jw,gengvec,Roca:2010tf}.  The most relevant mechanisms
 are the contact
term, Fig.~\ref{fig:vvdiagram1}a,
 and the $t,u$ channel exchange, Fig.~\ref{fig:vvdiagram1}b. 
The s-channel, Fig.~\ref{fig:vvdiagram1}c, is very small since it is
basically p-wave, and the box diagram, Fig.~\ref{fig:vvdiagram1}d, is
relevant only for the width of the generated resonance
\cite{Molina:2008jw,gengvec}.

In the present work only the $\rho\rho$ and $K^*\rho$ interactions are
needed. For the $\rho\rho$ interaction with spin 2,  the isospin 0
and 2 are possible. To the isospin 2 only the $\rho\rho$ channel contributes,
and to isospin 0 also $K^* \bar K^*$, $\phi\phi$, $\omega\omega$ and $\omega\phi$ contribute, but the
dominat coupling of the generated $f_2(1270)$ resonance is by far the $\rho\rho$ \cite{gengvec}.
This is one of the reasons 
why the  $f^*_2(1270)$ resonance is called a $\rho\rho$ molecule
or dynamically generated state from $\rho\rho$ interaction.
 For $K^*\rho$ interactions with spin 2 we have isospin 1/2
and 3/2. For isospin 1/2 the coupled channels needed are $K^*\rho$,
$K^*\omega$ and $K^*\phi$, and for isospin 3/2 only $\rho K^*$ is
present. However, in isospin 1/2, the coupling of the generated resonance $K^*_2(1430)$ to $K^*\rho$ is about five
times larger than to the other channels.
 
\begin{figure}[!t]
\begin{center}
\includegraphics[width=0.7\textwidth]{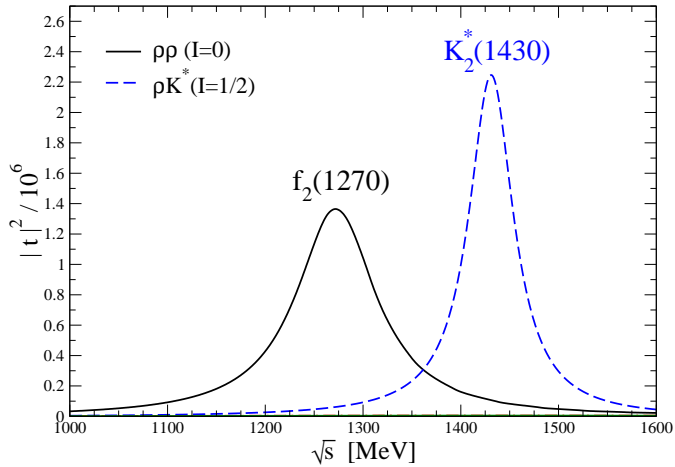}
\caption{Modulus squared of the $\rho\rho$ and $\rho K^*$ scattering amplitudes}
\label{fig:figT2elementary}
\end{center}
\end{figure}

The $\rho\rho$ and $K^*\rho$ amplitudes needed in the present work are shown in
Fig.~\ref{fig:figT2elementary}. The resonant structure for the $f_2(1270)$ and $K^*_2(1430)$
resonances are clearly visible in $t_{\rho\rho}^{(I=0)}$ and 
$t_{\rho K^*}^{(I=1/2)}$ respectively.
 The non-resonant amplitudes $t_{\rho\rho}^{(I=2)}$ and 
$t_{\rho K^*}^{(I=3/2)}$ are two orders of magnitude smaller compared to the resonant ones 
 and thus are not visible in the figure.

\section{Multi-body interaction\label{sec:3}}

The formalism to evaluate the interaction for more than two vector
mesons is similar to the one used in 
ref.~\cite{Roca:2010tf}. In ref.~\cite{Roca:2010tf} only $\rho$
mesons were involved. However, in the present work we have two different
species of particles, $\rho$ and $K^*$ resonances, which 
complicates a little bit the formalism. Therefore, in this section we
only show the novelties due to the fact of having two different
species and we refer the reader to ref.~\cite{Roca:2010tf} 
for the rest of
the formalism and for details.

We will illustrate the general process 
for the interaction of a generic particle  
$A$ interacting with a cluster $B$ made 
of two particles, $b_1$ and $b_2$. For instance, in order to generate
the  $K^*_3(1780)$ 
this would correspond to $A=K^*$, $B=f_2(1270)$ and $b_1=b_2=\rho$.

\begin{figure}[!t]
\begin{center}
\includegraphics[width=0.75\textwidth]{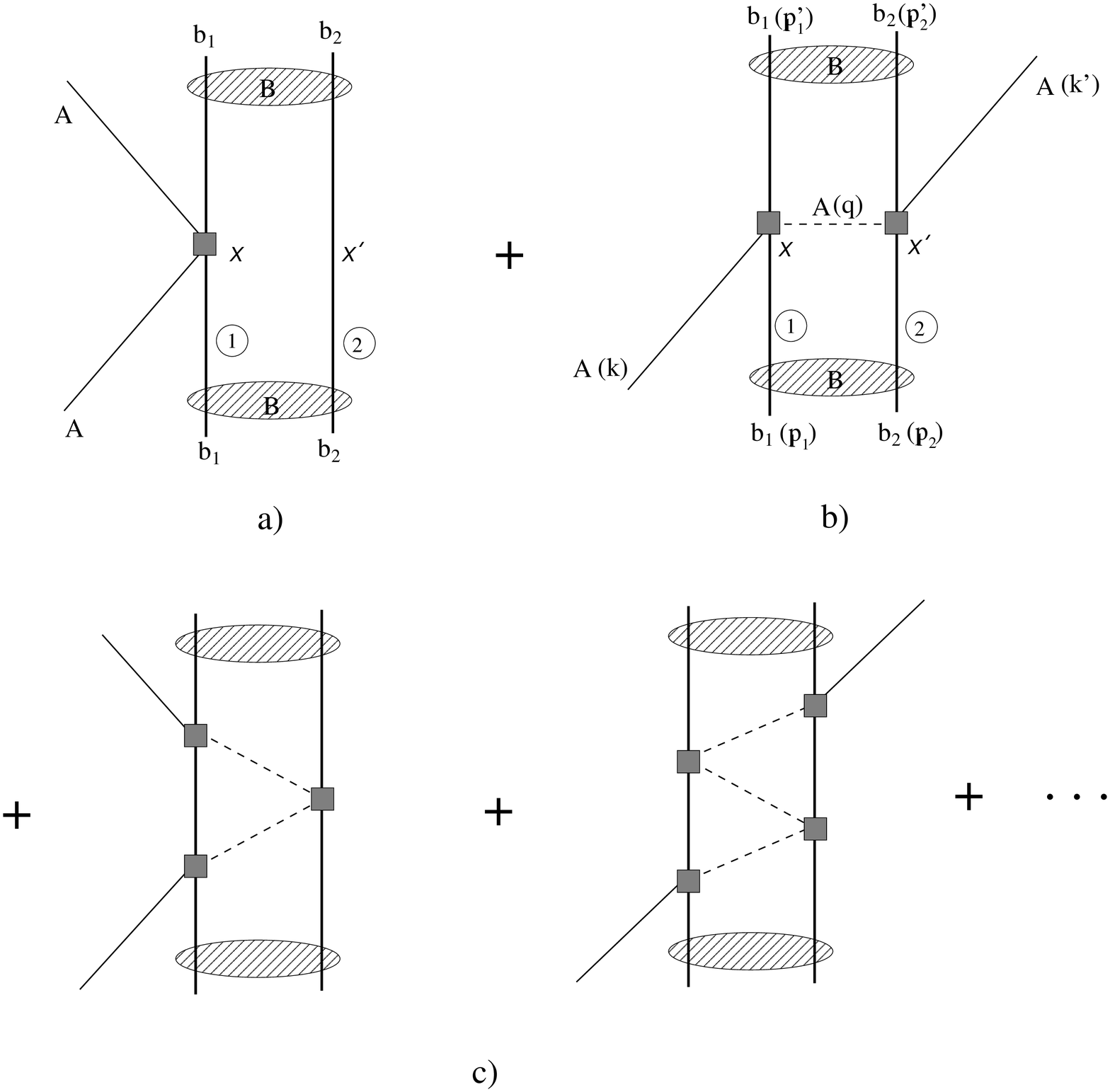}
\caption{Diagrammatic representation of the fixed center approximation
to the Faddeev equations for the interaction of a particle $A$ with a
particle $B$ made of a cluster of two particles, $b_1$ and $b_2$.
 Diagrams $a)$ and $b)$ represent
the single and double scattering contributions  respectively.}
\label{fig:Faddeev}
\end{center}
\end{figure}

The basic idea is to use the fixed center approximation of the Faddeev
 equations (FCA) 
 \cite{Faddeev:1960su,Chand:1962ec,Barrett:1999cw,Deloff:1999gc,Kamalov:2000iy},
  which are   written in terms of two partition functions $T_1$, $T_2$, which 
  sum up to the total scattering matrix, $T$, and read
\begin{eqnarray}
T_1&=&t_1+t_1G_0T_2\nonumber\\
T_2&=&t_2+t_2G_0T_1\nonumber\\
T&=&T_1+T_2
\label{eq:Faddeev}
\end{eqnarray}
where $T$ is the total scattering amplitude we are looking for, $T_i$
accounts for all the diagrams starting with the interaction of the
external particle $A$ with particle $b_i$ of the compound  system $B$ 
and
$G_0$ is the Green function for the exchange of a particle $A$ between
the $b_1$ and $b_2$ particles (dashed lines in Fig.~\ref{fig:Faddeev}).
A schematic representation for the FCA of Eq.~(\ref{eq:Faddeev})
 is depicted in Fig.~\ref{fig:Faddeev}.
The mechanism in Fig.~\ref{fig:Faddeev}a represents the
single-scattering contribution 
($t_1$ in Eq.~(\ref{eq:Faddeev}))
and Fig.~\ref{fig:Faddeev}b
the double-scattering mechanism (the next contribution, $t1=t1+t1G_0t_2)$). 
The addition of Fig.~\ref{fig:Faddeev}c represents the full resummation
of mechanisms to get the full $T_1$ partition function in the FCA. 
An analogous figure
starting with the particle $A$ interacting with $b_2$ would account for the 
$T_2$ amplitude.


For the evaluation of the two-body amplitudes, $t_1$ and $t_2$, in 
terms of the unitarized 
vector-vector amplitudes in isospin basis of Eq.~(\ref{eq:bethesalpeter}),
one has to take into account that the particles involved 
are in given isospin states.
We need first to consider the 
interaction
of a $K^*$ and a two-$\rho$ cluster.
The two $\rho$ mesons are in an isospin $I=0$ state,
\begin{equation}
|\rho\rho\rangle_{I=0}=-\frac{1}{\sqrt{3}}|\rho^+\rho^-+\rho^-\rho^++\rho^0\rho^0\rangle
=\frac{1}{\sqrt{3}}\Bigl(|(1,-1)\rangle+|(-1,1)\rangle-|(0,0)\rangle\Bigr)
\end{equation}
where the kets in the last member indicate the $I_z$ components of
 the $b_1$ and $b_2$ particles,~$|(I_z^{(b_1)},I_z^{(b_2)})\rangle$.
The external particle $A$ is a $K^*$ being in the state
 $|(I_z^{(A))}\rangle$ 
\begin{equation}
|K^*\rangle=|\frac{1}{2}\rangle~.
\end{equation}

The interaction in terms of the two body potentials $t_{Ab_1}$, $t_{Ab_2}$ 
can be written as
\begin{eqnarray}
T&=&\Bigl(\langle\frac{1}{2}|\otimes\frac{1}{\sqrt{3}}\langle(1,-1)+(-1,1)-(0,0)|\Bigr)(t_{Ab_1}+t_{Ab_2})\nonumber\\
&&\Bigl(|\frac{1}{2}\rangle\otimes\frac{1}{\sqrt{3}}|(1,-1)+(-1,1)-(0,0)\rangle\Bigr)\nonumber\\
&=&\frac{1}{3}\Bigl\langle((\frac{3}{2},\frac{3}{2}),-1)+\frac{1}{\sqrt{3}}((\frac{3}{2},-\frac{1}{2}),1)+\sqrt{\frac{2}{3}}((\frac{1}{2},-\frac{1}{2}),1)-((\frac{1}{2},\frac{1}{2}),0)\Bigl|t_{Ab_1}\nonumber\\
&&\Bigr|((\frac{3}{2},\frac{3}{2}),-1)+\frac{1}{\sqrt{3}}((\frac{3}{2},-\frac{1}{2}),1)+\sqrt{\frac{2}{3}}((\frac{1}{2},-\frac{1}{2}),1)-((\frac{1}{2},\frac{1}{2}),0)\Bigr\rangle\nonumber\\
&+&\frac{1}{3}\Bigl\langle\frac{1}{\sqrt{3}}((\frac{3}{2},-\frac{1}{2}),1)+\sqrt{\frac{2}{3}}((\frac{1}{2},-\frac{1}{2}),1)+((\frac{3}{2},\frac{3}{2}),-1)-((\frac{1}{2},\frac{1}{2}),0)\Bigl|t_{Ab_2}\nonumber\\
&&\Bigl|\frac{1}{\sqrt{3}}((\frac{3}{2},-\frac{1}{2}),1)+\sqrt{\frac{2}{3}}((\frac{1}{2},-\frac{1}{2}),1)+((\frac{3}{2},\frac{3}{2}),-1)-((\frac{1}{2},\frac{1}{2}),0)\Bigr\rangle
\label{eq:larga1}
\end{eqnarray}
where the notation followed in the last term for the states is $\langle(I^{\rm total},~I^{\rm total}_z,~I^k_z)|t_{ij}|\rangle$, where $I^{\rm total}$ means the total isospin of the $ij$ system and $k\ne i,~j$ (the spectator particle).

This leads to the following amplitude 
for the single scattering contribution,
\begin{equation}
t_{\rho K^*}=\frac{1}{9}\Bigr(4t_{\rho K^*}^{\rm (I=\frac{3}{2})}+5t_{\rho K^*}^{\rm (I=\frac{1}{2})}\Bigl)~.
\label{eq:trhoKsa}
\end{equation}


In the evaluation of some of the amplitudes in the present work, 
we will also need to consider the interaction of a $\rho$ meson 
and a  $K^*_2$ cluster. But in this case, the
$\rho K^*$ amplitude needed is different than Eq.~(\ref{eq:trhoKsa})
and the $\rho\rho$ amplitude is different than in
 ref.~\cite{Roca:2010tf} since now the cluster
 is a $K^*\rho$ in isospin $I=1/2$:
 
\begin{equation}
|\rho K^*_2\rangle_{I=\frac{1}{2},I_z=\frac{1}{2}}=
 \sqrt{\frac{2}{3}}  \Bigl(|(1,-\frac{1}{2})\rangle
-\frac{1}{\sqrt{3}} |(0,\frac{1}{2})\rangle\Bigr)
\label{eq:ketrhoks2}
\end{equation}
where we have taken the $I_z=1/2$ for convenience and, 
in analogy to the nomenclature used in the equations above, 
the bracket represents $(I_z^{\rho},I_z^{K^*_2})$.
Therefore the $\rho K$ inside the $K^*_2$ can be in $I_z=-1/2$ and
$I_z=+1/2$, with the states given by
\ba
|\rho K^*\rangle_{I=\frac{1}{2},I_z=-\frac{1}{2}}&=&
 \frac{1}{\sqrt{3}}   \Bigl(|(0,-\frac{1}{2})\rangle
- \sqrt{\frac{2}{3}}|(-1,\frac{1}{2})\rangle\Bigr)\nonumber\\
|\rho K^*\rangle_{I=\frac{1}{2},I_z=+\frac{1}{2}}&=&
 \sqrt{\frac{2}{3}}  \Bigl(|(1,-\frac{1}{2})\rangle
-\frac{1}{\sqrt{3}} |(0,\frac{1}{2})\rangle\Bigr)
\label{eq:ketrhoks}
\ea
where now the bracket represents $(I_z^{\rho},I_z^{K^*})$.
Combining Eqs.~(\ref{eq:ketrhoks2}) and (\ref{eq:ketrhoks})
and proceeding analogously to Eq.~(\ref{eq:larga1}) we get, after a bit
of algebra, that the two-body amplitudes needed in the evaluation of
 the
$\rho K^*_2$ interaction are
\ba
t_{\rho\rho}&=&
\frac{2}{3}t_{\rho\rho}^{(I=0)}\nonumber\\
t_{\rho K^*}&=&\frac{1}{9}\Bigr(8t_{\rho K^*}^{\rm (I=\frac{3}{2})}
+t_{\rho K^*}^{\rm (I=\frac{1}{2})}\Bigl)~.
\label{eq:10}
\ea
Note that, as mentioned above when explaining Fig.~\ref{fig:figT2elementary}, 
the $t_{\rho K^*}^{\rm (I=\frac{3}{2})}$ term is
negligible compare to $t_{\rho K^*}^{\rm (I=\frac{1}{2})}$ in spite of being multiplied by
a factor 8 in the previous equation.
The $t_{\rho\rho}^{\rm (I=0)}$ of Eq.~(\ref{eq:10}) is evaluated with the unitary normalization for two identical $\rho$ mesons (extra factor $1/\sqrt{2}$) \cite{Molina:2008jw}.

On the other hand, 
it is worth noting that the
 argument of the function $T(s)$ in the FCA
 is the total invariant mass energy
$s$, while the argument of $t_1$ and $t_2$ are $s_1$ and $s_2$, where
$s_i (i=1,~2)$ is the invariant mass of the interaction particle $A$ and
the particle $b_i$ of the $B$ molecule and is given by

\begin{equation}
s_i=m_A^2+m_{b_i}^2+\frac{1}{2 m_B^2}(s-m_A^2-m_B^2)
(m_B^2+m_{b_i}^2-m_{b_{j\ne i}}^2),
\end{equation}
where $m_{A(B)}$ is the mass of the $A(B)$ 
system and $m_{b_i}$ is the mass of each building block
 of the $B$ molecule.

In order to obtain the solution of the FCA, Eq.~(\ref{eq:Faddeev}),
in terms of the two-body amplitudes $t_{Ab_i}$ and with the proper
normalization,
let us first consider the wavefunctions of the incident and outgoing 
$A$ particle
being plane waves normalized inside a box of volume $\cal
V$. Then, following the same
 calculation as in ref.~\cite{Roca:2010tf}, the
$S$-matrix for the single-scattering process of
 Fig.~\ref{fig:Faddeev}a is written as

\ba
S^{(1)}&=&-it_{Ab_1} \frac{1}{{\cal V}^2}
\frac{1}{\sqrt{2\omega_{p_1}}}
\frac{1}{\sqrt{2\omega_{p'_1}}}
\frac{1}{\sqrt{2\omega_k}}
\frac{1}{\sqrt{2\omega_{k'}}}\nonumber\\
&&\times(2\pi)^4\,\delta(k+k_{B}-k'-k'_{B}).
\label{eq:Ssingle}
\ea
where the momenta are defined in Fig.~\ref{fig:Faddeev}b,
 $\omega_p$ represents the on-shell energy of the corresponding
particle with momentum $p$ and $k_B$ ($k'_B$)
represents the total momentum of the
initial (final) cluster $B$.


The next contribution is the double scattering mechanism,
which corresponds to Fig.~\ref{fig:Faddeev}b.
Proceeding similarly to ref.~\cite{Roca:2010tf} the $S$-matrix for this
contribution
takes the form
\ba
S^{(2)}&=&-i(2\pi)^4 \delta(k+K_{B}-k'-K'_{B})\frac{1}{{\cal V}^2}
\frac{1}{\sqrt{2\omega_k}} 
\frac{1}{\sqrt{2\omega_{k'}}}
\frac{1}{\sqrt{2\omega_{p_1}}}
\frac{1}{\sqrt{2\omega_{p'_1}}}
\frac{1}{\sqrt{2\omega_{p_2}}} 
\frac{1}{\sqrt{2\omega_{p'_2}}}\nn\\
&&\times\int \frac{d^3q}{(2\pi)^3} 
F(q,B)
\frac{1}{{q^0}^2-\vec{q}\,^2-m_A^2+i\epsilon} t_{Ab_1} t_{Ab_2}.
\label{eq:finalS2}
\ea
where $F(q,B)$ is the form factor of the particle $B$, which represents
essentially the Fourier transform of its wave function. Its
derivation and further interpretation 
can be found in ref.~\cite{Roca:2010tf} and its mathematical 
expression is given
below,
in Eq.~(\ref{eq:ffact}).

On the other hand, the $S$-matrix of a general
 $AB$ interaction can be written
as

\ba
S&=&-i t_{AB}(2\pi)^4 \delta(k+K_{B}-k'-K'_{B})\frac{1}{{\cal V}^2}\nn\\
&&\times\frac{1}{\sqrt{2 \omega_k}} 
\frac{1}{\sqrt{2 \omega_{k'}}}
\frac{1}{\sqrt{2 \omega_{k_B}}}
\frac{1}{\sqrt{2 \omega_{k'_B}}}.
\label{eq:finalS2AB}
\ea

Comparing Eq.~(\ref{eq:finalS2}) and (\ref{eq:finalS2AB}), we can
deduce the form that the  
FCA equations (\ref{eq:Faddeev})  take in our case
\begin{eqnarray}
T_{Ab_1}&=& c_1 + c_1 \widetilde{G}_0 T_{Ab_2}  \nonumber\\
T_{Ab_2}&=& c_2 + c_2 \widetilde{G}_0 T_{Ab_1}\nonumber\\
T&=&T_{Ab_1}+T_{Ab_2}
\label{eq:Faddeevsystem}
\end{eqnarray}
with
\be c_i=  \frac{m_B}{m_{b_i}}t_{A b_i}(s_i),
\ee
where we have taken 
$\omega_{p_i}\simeq m_i$, and
\footnote{Note the factor 1/2 different in Eq.~(\ref{eq:G0}) with
respect to Eq.~(41) of ref.~\cite{Roca:2010tf} due to the new, more general, formulation.}
\ba
 \widetilde G_0(s;A,B)=\frac{1}{2 m_B}
\int\frac{d^3q}{(2\pi)^3} 
F(q;B)
\frac{1}{{q^0(s;A,B)}^2-\vec{q}\,^2-m_A^2+i\epsilon}.
\label{eq:G0}
\ea
In Eq.~(\ref{eq:G0}), $q^0$ is 
\be
q^0(s;A,B)=\frac{s-m_A^2-m_B^2}{2m_B}
\ee
and the form factor takes de form 
\ba
F(q,B)=\frac{1}{ {\cal N}}
\int_{\substack{p<\Lambda'\\|\vec p-\vec q|<\Lambda'}}
d^3p\,
\frac{1}{m_B-\sqrt{\vec{p}\,^2+m_{b_1}^2}
-\sqrt{\vec{p}\,^2+m_{b_2}^2}}\nonumber\\
\times\frac{1}{m_B-\sqrt{|\vec p-\vec q|^2+m_{b_1}^2}
            -\sqrt{|\vec p-\vec q|^2+m_{b_2}^2}},
\label{eq:ffact}
\ea 
\ba
{\cal N}=
\int_{p<\Lambda'}d^3p
\frac{1}{\left( m_B-\sqrt{\vec{p}\,^2+m_{b_1}^2}
-\sqrt{\vec{p}\,^2+m_{b_2}^2}\right)^2}.
\label{eq:ffactnorm}
\ea

In Eq.~(\ref{eq:ffact}), $\Lambda'$ represents a cutoff 
which has a similar physical meaning \cite{Roca:2010tf}
as the three-momentum cutoff of the
loop function of  Eq.~(\ref{eq:Gsharp}). For that reason we take for
$B=f_2$ and $B=K^*_2$ the same values for $\Lambda'$
as for the cutoffs mentioned below 
Eq.~(\ref{eq:Gsharp}). For $B=f_4$  and $B=K^*_4$, which we also
need in the present work, in principle the $\Lambda'$ could be
different since the typical maximum momentum could be
 different. We have
checked that changing reasonably the value of $\Lambda'$ for these
cases as done in ref.~\cite{Roca:2010tf} the qualitative conclusions
do not change and the quantitative differences are small.

The system of equations~(\ref{eq:Faddeevsystem}) can be
algebraically solved and it gives for the final multi-body scattering
amplitude

\be
T_{AB}=T_{Ab_1}+T_{Ab_2}=\frac{c_1+c_2+2c_1c_2\widetilde G_0}
{1-c_1c_2 \widetilde G_0^2}.
\ee

\section{Results}
\label{sec:results}

The interactions that we consider in the present work, summarized in
table~\ref{tab:clusters},
are the following.
For three particles: $A=K^*$, $B=f_2=(b_1=\rho,b_2=\rho)$ and $A=\rho$,
$B=K^*_2,(b_1=\rho,b_2=K^*)$. For four particles:  $A=f_2$, $B=K^*_2=(b_1=\rho,b_2=K^*)$.
For five particles:  $A=K^*$, $B=f_4=(b_1=f_2,b_2=f_2)$ and
 $A=\rho$, $B=K^*_4=(b_1=f_2,b_2=K^*_2)$.
 For six particles:  $A=K^*_2$, $B=f_4=(b_1=f_2,b_2=f_2)$ and
 $A=f_2$, $B=K^*_4=(b_1=f_2,b_2=K^*_2)$.
(By the number of particles above we mean the number of elementary vector mesons
 if each cluster or molecule is broken down
into its vector meson constituents).

\begin{table}[h]
\begin{center}
\begin{tabular}{|c|c|c|} 
\hline
& $A$ & $B$ $(b_1 b_2)$  \\ \hline
 two-body &  $\rho$ & $K^*$   \\ \hline
 three-body &\begin{tabular}{c} $K^*$\\ $\rho$\end{tabular}  & 
           \begin{tabular}{l} $f_2$ $(\rho\rho)$ \\ $K^*_2$ $(\rho K^*)$\end{tabular}
	   \\ \hline
 four-body & $f_2$ & $K^*_2$ $(\rho K^*)$  \\ \hline
 five-body & \begin{tabular}{l} $K^*$\\ $\rho$\end{tabular}  & 
           \begin{tabular}{l} $f_4$ $(f_2 f_2)$\\ $K^*_4$ $(f_2 K^*_2)$\end{tabular}  \\ \hline
 six-body &   \begin{tabular}{l} $K^*_2$\\ $f_2$\end{tabular} & 
           \begin{tabular}{l} $f_4$ $(f_2 f_2)$ \\ $K^*_4$ $(f_2 K^*_2)$\end{tabular} \\ \hline
 \end{tabular}
\end{center}
\caption{Clusters considered in the evaluation of the multi-body interactions. 
(See Fig.~\ref{fig:Faddeev} for nomenclature).}
\label{tab:clusters}
\end{table}

\begin{figure*}
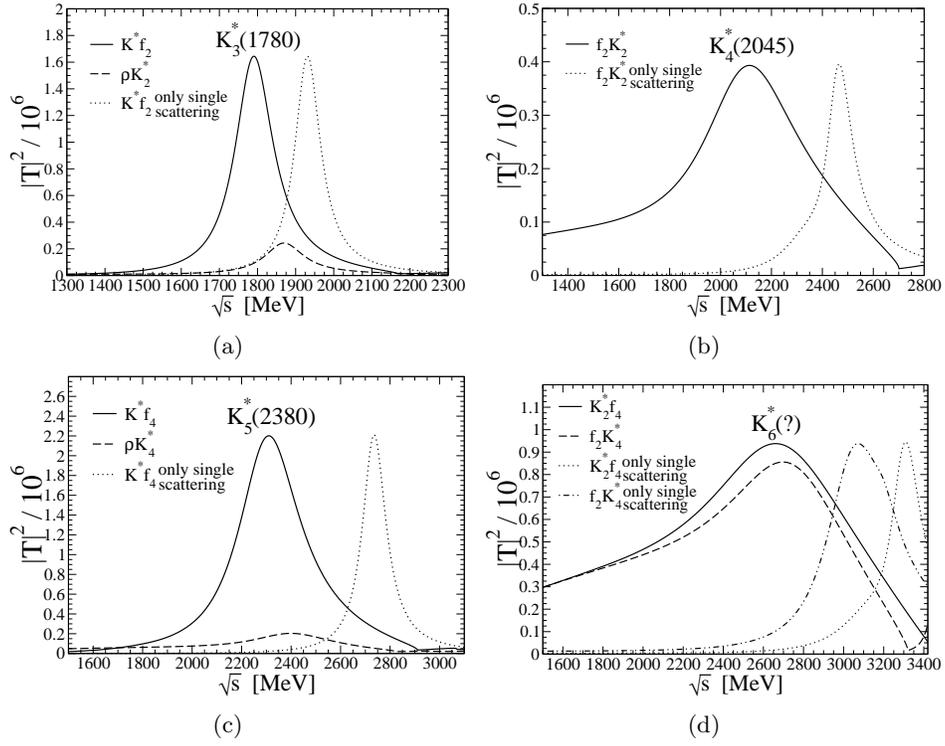

     \centering
      \subfigure[]{
          \label{fig:T2_Ks3}
          \includegraphics[width=.48\linewidth]{figure4a.eps}}
      \subfigure[]{
          \label{fig:T2_Ks4}
          \includegraphics[width=.48\linewidth]{figure4b.eps}}
       \subfigure[]{
          \label{fig:T2_Ks5}
          \includegraphics[width=.48\linewidth]{figure4c.eps}}
      \subfigure[]{
          \label{fig:T2_Ks6}
          \includegraphics[width=.48\linewidth]{figure4d.eps}}
   \caption{Modulus squared of the unitarized $K^*$--multi-$\rho$ amplitudes.
   The dotted and dashed-dotted lines have been normalized to
   the peak of the solid line for the sake of comparison 
   of the position 
   of the maxima}
     \label{fig:T2s}
\end{figure*}

In Fig.~\ref{fig:T2s} we show the modulus squared of the different multy-body scattering 
amplitudes.
The dotted and dashed-dotted lines represent the calculation considering only the single scattering
mechanism and are normalized to the peak of the solid lines, which represent the full amplitudes,
 for the sake of comparison
of the position of the maxima.
The resonant structure of the amplitudes is clearly evident in the plot,
which can be associated to the resonances
labeled in the figures with masses given by the position of the maxima.

In table~\ref{tab:masses} the values of the masses of our generated
systems are shown in comparison with the experimental
values at the PDG \cite{Amsler:2008zzb}. Note that the $K^*_6$ resonance is not quoted in the
PDG, therefore our results on this resonance are genuine predictions of our model. Furthermore, it is also worth mentioning
that there is only one single experiment \cite{Aston:1987ir} reporting on the existence of the  $K^*_5(2380)$ and thus in
the PDG it is quoted as ``needs confirmation''.

\begin{table}[h]
\begin{center}
\begin{tabular}{|c|c|c|c|c|} 
\hline
  \begin{tabular}{c} generated \\ resonance \end{tabular} &amplitude & mass, PDG \cite{Amsler:2008zzb} 
& \begin{tabular}{c}mass \\ only single scatt. \end{tabular} & \begin{tabular}{c}mass \\ full model \end{tabular}  \\ \hline
$K^*_2(1430)$ & $\rho K^*$            & $1429\pm 1.4$ & $-$    & 1430	 \\ \hline
$K^*_3(1780)$ & $K^*f_2$              & $1776\pm   7$ & 1930 &1790   \\ \hline
$K^*_4(2045)$ & $f_2K^*_2$            & $2045\pm   9$ & 2466     &2114   \\ \hline
$K^*_5(2380)$ & $K^*f_4$              & $2382\pm 14\pm 19$ & 2736 &2310  \\  \hline
$K^*_6$   &  $K^*_2f_4$-$f_2K^*_4$ &   $-$  & 3073-3310 &2661-2698    \\ \hline
 \end{tabular}
\end{center}
\caption{Results for the masses of the dynamically generated states. (All units are MeV)}
\label{tab:masses}
\end{table}
The second column of table~\ref{tab:masses} indicates the dominant amplitude among those considered to get the resonance from
which we take the position of the maximum. That amplitude is the one chosen to get the mass quoted in the table, from the
position of the maximum. For instance, for the  $K_3^*$ we see in  Fig.~\ref{fig:T2_Ks3} that the $K^*f_2$ squared amplitude
is about seven times bigger than for $\rho K^*_2$.  For the $K^*_6$ both channels considered have almost
equivalent strength, therefore both are considered in order to get the predicted mass of this resonance.

With only the single scattering mechanism, qualitative bumps for the resonances are produced but with masses that differ from
the experimental values in about $200-400\mev$. The consideration of the multiple scattering processes
in the FCA
 improves drastically the
agreement with the experimental values of the masses. We have checked (not shown in
Fig.~\ref{fig:T2s}) that only up to double scattering, Fig.~\ref{fig:Faddeev}a and b, the position of the masses do not
improve significantly  from considering only the single scattering. Therefore,
 the full resummation of  Fig.~\ref{fig:Faddeev}c is
crucial in order to get the full amplitude. This is a clear manifestation of the non-perturbative character of the  Faddeev
equations~(\ref{eq:Faddeev}).

The agreement in the masses of our full model with the experimental values is remarkable, specially considering the large widths of these resonances. 
It is worth stressing the simplicity of our approach 
and the absence of
parameters fitted in the model once the three-momentum cutoffs
 are chosen to get the right mass of the $f_2(1270)$ and $K^*_2(1430)$ resonances in the way explained below 
Eq.~(\ref{eq:Gsharp}).
Obtaining the widths of the generated resonances from the amplitudes is a more involved issue \cite{Roca:2010tf}.
The widths of the bumps in the modulus squared of the amplitudes could in principle be associated to the resonance widths if they
 were Breit-Wigner like shapes,
which is not the case. It is worth stressing that our procedure produces the full amplitude,
not only the resonant contribution. This means that the amplitudes
contain also possible non-resonant or background contributions. 
The difference from Breit-Wigner shapes manifests the strong background produced by the non-linear dynamics of the non-perturbative
 calculation carried out in the present
work. One has to consider that in a real experiment much of the strength of the amplitude would be associated to a background, which would
reduce the width assigned to the resonance with respect the visual 
or apparent one. Taking this into account we can estimate semi-qualitatively the widths of the
generated resonances as
120, 300, 300 and $600\mev$ for $K^*_3$, $K^*_4$, $K^*_5$ 
and $K^*_6$ to be compared to the experimental values \cite{Amsler:2008zzb} $159\pm 21$, $198\pm 30$, $178\pm37\pm 32$ and {\it undetected} respectively.
The width is also more sensitive to the value used in the cutoff of Eq.~(\ref{eq:ffact}), (see the discussion
below Eq.~(\ref{eq:ffactnorm})). So the estimation of
the widths must be considered only as qualitative. Regarding the $K^*_6$, it has not yet been discovered and thus
we  predict  from our results  a resonance  $I(J^P)=1/2(6^+)$ with a mass 
about $2650-2750\mev$ and width 
around several hundreds MeV.
Certainly, this large width does not make its possible identification easy.

\section{Conclusions}
\label{sec:concl}

We show in the present work that the resonances $K^*_2(1430)$, $K^*_3(1780)$, $K^*_4(2045)$, $K^*_5(2380)$ 
and $K^*_6$ are basically states made of one  $K^*(892)$ and an increasing number of $\rho(770)$ resonances.
The point is that the interaction between two vector mesons with spins aligned is very strong and, therefore, they tend to cluster. 
In particular, in refs.~\cite{Molina:2008jw,gengvec} the $\rho\rho$ and $\rho K^*$
interaction was shown to be strong enough to bind 
the particles
forming the $f_2(1270)$ and the $K^*_2(1430)$ resonances respectively. 
These vector-vector interactions were obtained by implementing the techniques of the chiral unitary approach with the only input of the lowest
order vector-vector potential obtained from suitable hidden gauge symmetry lagrangians.

The multi-body amplitude in terms of the elementary vector-vector interactions is evaluated by solving the Faddeev equation in the fixed center approximation, which
considers the interaction between three particles where two of them are clustered forming a resonance.

In the amplitudes obtained, there are significant bumps that can be associated to these specific resonances and we can obtain the mass from the position at the
maxima. We get remarkable agreement in the mass of the resonances with the experimental values and give a prediction for the undiscovered state $K^*_6$. It is also
worth stressing the conceptual simplicity of the model and the non-inclusion of free parameters in the present work, once two regularization parameters are fixed
in refs.~\cite{Molina:2008jw,gengvec} to get the proper mass of the $f_2$ and $K^*_2$ resonances.
It is clear that more states, like quark-antiquark, other meson-meson, several mesons, etc, can contribute to
the wave functions of these resonances, but the fact that we get the resonances and in good agreement with
experiment indicates that the many-vector-mesons components are dominant in the building up of the resonances
studied in the present work. 

 Further experimental studies, specially about the
not-yet discovered $K^*_6$ and the poorly studied $K^*_5(2380)$,
 would be welcome to clarify the issue on the nature 
of these resonances. The success of the techniques used in the present work 
should also spur their future applications to other systems.

\section*{Acknowledgments}
We thank L.~Geng and R.~Molina for providing us the vector-vector amplitudes.
This work is partly supported by DGICYT contracts  FIS2006-03438,
FPA2007-62777, the Fundaci\'on S\'eneca grant 11871/PI/09 and
the EU Integrated Infrastructure Initiative Hadron Physics
Project  under Grant Agreement n.227431.

\end{document}